\documentclass[12pt,journal,compsoc]{IEEEtran}
\usepackage{amsmath,amsfonts}
\usepackage{algorithmic}
\usepackage{array}
\usepackage[caption=false,font=normalsize,labelfont=sf,textfont=sf]{subfig}
\usepackage{textcomp}
\usepackage{stfloats}
\usepackage{url}
\usepackage{verbatim}
\usepackage{graphicx}
\def\BibTeX{{\rm B\kern-.05em{\sc i\kern-.025em b}\kern-.08em
    T\kern-.1667em\lower.7ex\hbox{E}\kern-.125emX}}
\usepackage{balance}
\usepackage{graphicx,psfrag,pslatex,color,eepic,verbatim,comment}

\newcommand{\refs}[1]{Section~\ref{#1}}

\newcommand{\reff}[1]{Fig.~\ref{#1}}

\begin{document}
\title{Multiple Access in Constellation Domain by Non-Coherent Massive MIMO}
\author{{Victor Monzon Baeza} \\[2.5mm]
    \textit{Interdisciplinary Centre for Security Reliability and Trust (SnT), SIGCOM Group,\\ University of Luxembourg, Luxembourg}\\
    {{\tt victor.monzon@uni.lu}}\\[6mm]}

\markboth{~Vol.~XX, No.~X, October~2022}%
{How to Use the IEEEtran \LaTeX \ Templates}

\maketitle

\date{} 

\begin{abstract}
Multiple access is the base for increasing the capacity in multi-user communication networks. However, the growing demand for higher data rates and the number of users who requires communication services has led to the scarcity of orthogonal resources in current wireless communications. On the other hand, integrating the satellite within terrestrial networks as an initiative of 3GPP since its Release 15 entails the need for new forms of multiple access between terrestrial and non-terrestrial users. This paper studies constellation schemes as a new domain to enhance the state-of-the-art multiple-access techniques for future communication technologies employing non-coherent communications with massive MIMO. In addition, we propose a hybrid model between the classic access methods such as Time Division Multiple Access (TDMA) or Frequency Division Multiple Access (FDMA), the emerging models of non-orthogonal multiple access (NOMA) and the proposed domain of the constellation based on non-coherent massive multiple-input multiple-output (MIMO) schemes. This model is discussed for different scenarios in satellite communications that help increase the system's capacity and avoid interference between terrestrial and non-terrestrial users.
\end{abstract}


\section{Introduction}
\label{intro}

The explosive use of communication technologies claiming indigent amounts of data has led to a shortage in radio resources for sharing them across multiple users in a cell. Traditional radio resources such as time, frequency, or codes have allowed multiple users to access the services provided simultaneously. These resources are used as multiple orthogonal accesses until the third generation of wireless communications (1G-3G). Orthogonal Frequency Division Multiplexing (OFDM) was employed for the fourth generation (4G) \cite{1}. The scarcity of more orthogonal resources led the research community to explore alternatives to its non-orthogonal counterparts, giving rise to Non-Orthogonal Multiple Access (NOMA), evaluated in \cite{2} and proposed for the fifth generation (5G) in \cite{3}. 

The need for global coverage in mobile networks has led 3GPP to propose the integration of the space segment or non-terrestrial networks (NTN) with terrestrial networks. A complete network in which the satellite is one more member of the terrestrial system transparently is the objective of the future sixth generation (6G). This need has opened a line of research for new multiple-access techniques. The problem of the scarcity of resources to simultaneously access the services is accentuated when we integrate more elements in the network, together with a new issue; the management of interference between users of the terrestrial network with those who access the non-terrestrial network. Advances in access multiple considering TDMA and NOMA have been proposed in a framework for satellites in \cite{5}. However, the authors do not consider the integration with terrestrial networks. 
In addition, the 5G physical layer includes massive MIMO technology to increase spectral and energy efficiency \cite{6}. The problem with this technology is the enormous amount of information that must be processed to estimate many channels. Therefore, to overcome this problem, non-coherent (NC) techniques are emerging as a potential candidate for the future of communications in 6G networks \cite{7}. The union of the Non-Coherent schemes with massive MIMO solves the problems of the absence of channel information, reaching the performance of its coherent counterpart regarding channel capacity and bit error rate (BER). 

NOMA research has generated many works in recent years, giving rise to new domains for sharing radio resources simultaneously by users. For example, Spatial Division Multiple Access (SDMA), Rate Splitting  Multiple Access (RSMA), or Pattern Division Multiple Access (PDMA) are some of the new domains raised for multiple access within the NOMA group. This multitude of proposals has led the authors in \cite{8} to analyse the suitability of NOMA with MIMO-based schemes. At the same time, the authors in \cite{9} analyse the performance of multicarrier waveforms with new strategies for multiple access. 
Given this situation, new domains to simultaneously share radio resources between users should be proposed, considering the new elements integrated into 6G networks beyond the current ones offered by NOMA or classic techniques to evolve communications networks.  

The main contribution of this article is to propose the constellation schemes as a new multiple-access domain in the satellite communication networks to consider the integration of the satellite with the terrestrial networks in 6G. This new domain is based on the NC schemes union with massive MIMO to enhance the state-of-the-art multiple-access techniques empowering their performance. In addition, we propose use cases for different scenarios where the satellite is included under the constellation domain. 

The rest of the article is organized as follows: \refs{sec:background}  summarizes the contributions and advances in communications not consistent with massive MIMO. \refs{sec:model} presents the multiple access model proposed for future communication systems. \refs{sec:hybrid} proposes a hybrid scheme for multiple access. \refs{sec:op} proposes the scenarios in which the satellite is integrated as another member of the terrestrial networks. \refs{sec:ch} lists the challenges that emerge for future research. Finally, \refs{sec:conclusions} highlights the conclusions.


\section{Non-Coherent Background}
\label{sec:background}

Non-coherent techniques have always attracted a lot of interest due to their ability to avoid channel estimation, which saves a lot of resources for communication receivers. However, their well-known loss of 3 dB in performance compared to their coherent counterpart \cite{16} and the lack of hardware capacity to carry them out in practical systems have always kept them in the background. With the appearance of massive MIMO, they have regained their place since their performance has improved, achieving that achieved with coherent schemes. 
 
The union between massive MIMO and non-coherent schemes has focused on the physical layer's constellation. All these proposals consider a system with a single antenna transmitter and a single receiver with a large number of antennas. Users access the system using different constellation schemes between them. The designs proposed in the literature follow two main research currents: 

\begin{enumerate}
    \item Energy-based schemes.
    
    \item Phase-based schemes.
\end{enumerate}

As for the first group, reference examples are shown in \cite{10}. The authors in \cite{10} propose a non-coherent scheme empowered by massive MIMO in which a transmitter modulates information in the signal's amplitude and a receiver that measures the average received energy across the antennas. The constellation scheme employed is Amplitude Shift Keying (ASK) which has as many energy levels as the number of users in the system. This model is characterized by the simplicity of the circuit design and energy efficiency. However, the effect of atmospheric attenuation on the signal amplitude makes it undesirable to be able to integrate the satellite into 6G networks.   
 
On the other hand, we find more designs related to the signal phase to get users to access the medium. 
The authors in \cite{11} presented a design based on Phase Shift Keying (PSK) for a multi-user massive MIMO receiver that requires the users' angular direction. They compare several strategies for the implementation, such as one-bit Sigma-Delta quantization. Unfortunately, the stability of these systems is still an open issue. 

The dissemination in \cite{Thesis} presented the basis for different M-ary differential phase-shift keying (DPSK)-based schemes, including non-coherent schemes with massive MIMO.
Later works \cite{12,13,14} incorporated differential coding to improve the system's performance; remarkably, the receiver's complexity was reduced thanks to the massive MIMO detection process. 
In \cite{12}, a new constellation is designed for the multi-user non-coherent massive MIMO system based on DPSK. In addition, to reduce the number of antennas required for the receiver, the authors introduce a bit-interleaved coded modulation-iterative decoding scheme (BICM-ID). This proposal shows an overall improvement of two orders of magnitude, dramatically reducing the required number of antennas. 
In \cite{13}, the previous design is analysed over time-varying channels defining the number of receive antennas required in hostile media for attaining a similar performance to that achieved for stationary channels. Additionally, the maximum achievable rate is analysed. The results show that the non-coherent multiplexing empowered by massive MIMO is very robust to the channel’s temporal correlation effects. Therefore, this system may be ideal for future mobile communications networks over high-speed channels, for example, trains or aeronautical communications. In addition, we must consider that to increase the bandwidth in future communication systems, the frequency band in millimeters and THz are potential candidates. In these bands, the channel becomes more directive, highlighting the component of direct vision (LOS). In \cite{single}, the authors have studied the effect of the LOS component in NC schemes with massive MIMO based on DPSK for the case of a single user, while in \cite{rice} for multiple users. Another consideration for non-coherent schemes is heterogeneous channel conditions like those discussed in \cite{grouping} where LOS and non-LOS components are present. 
The authors in \cite{14} go one step further by adding SDMA to the DPSK design. In this scheme, each user firstly maps its bit stream to sparse code-words and then performs DPSK modulation on the non-zero dimensions. 

\section{Multiple Access Model in Constellation Domain}
\label{sec:model}

\begin{figure*}[!ht]
\centering
\includegraphics[width=1.75\columnwidth]{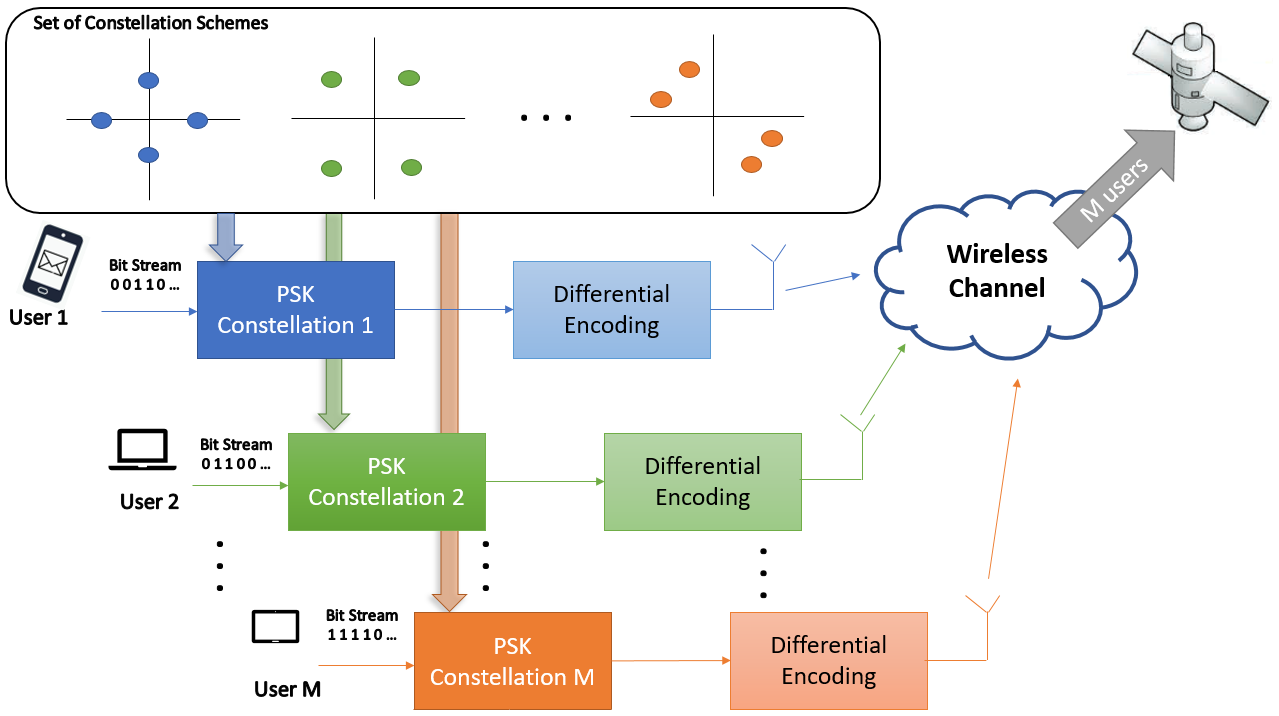}
\caption{Multiple access based on constellation schemes.}
\label{system}
\end{figure*}

The new domain proposed to share the resources by the users is the constellation used to modulate the signals in the physical layer. Since our objective is to perform multiple access between terrestrial and non-terrestrial users, we consider the base model of the existing for non-coherent those based on phase, specifically with differential encoding as shown in \cite{12,13}. This new access method uses a different constellation scheme for each user from a possible set, as shown in \reff{system}. The base model for each constellation is taken from a PSK scheme where each user places their symbols in a different position along a unit circle based on two main criteria:

\begin{itemize}
    \item \textbf{Criteria for Unequal Error Protection (UEP)}: each user will achieve a different performance level. In this case, the symbols are placed at different distances from each other.  
    
    \item \textbf{Criteria for Equal Error Protection (EEP)}: this criterion is based on placing all symbols of a given user at equal distances and keeping the same distance between any two symbols of all the users. This way, all users achieve the same performance level.
\end{itemize} 

If users want different quality of service (QoS), it is recommended to use UEP. An EEP should be used if all users require the same service benefits. For example, to consider integrating terrestrial users with non-terrestrial users in 6G, a UEP scheme is recommended. Instead, users should use the EEP criterion within the same network (terrestrial or NTN).
  
Each scheme employed by a user will be called individual constellation. After performing a modulation using the individual constellation, each user will apply a differential coding before transmitting the signal, as shown in \reff{system}.
On the receiving side, all the signals from all the users that access simultaneously are collected in each of the antennas that make up the massive MIMO system. For example, a regenerative payload performs the user separation as shown \reff{system}. The superposition of all the signals on the receiving side forms a scheme called joint constellation. The formation process of the joint constellation is detailed in \cite{12}. This superposition of signal in joint constellation means the additive combination of all individual constellations as described in \cite{12}. For this reason, the design of the individual constellations for each user must be meticulously designed previously in such a way that the combination of all the possible symbols does not give two equal joint symbols in the joint constellation; otherwise, the separation of the users in the receiver (payload) does not it will be possible. 

\begin{figure}[!ht]
\centering
\includegraphics[width=1\columnwidth]{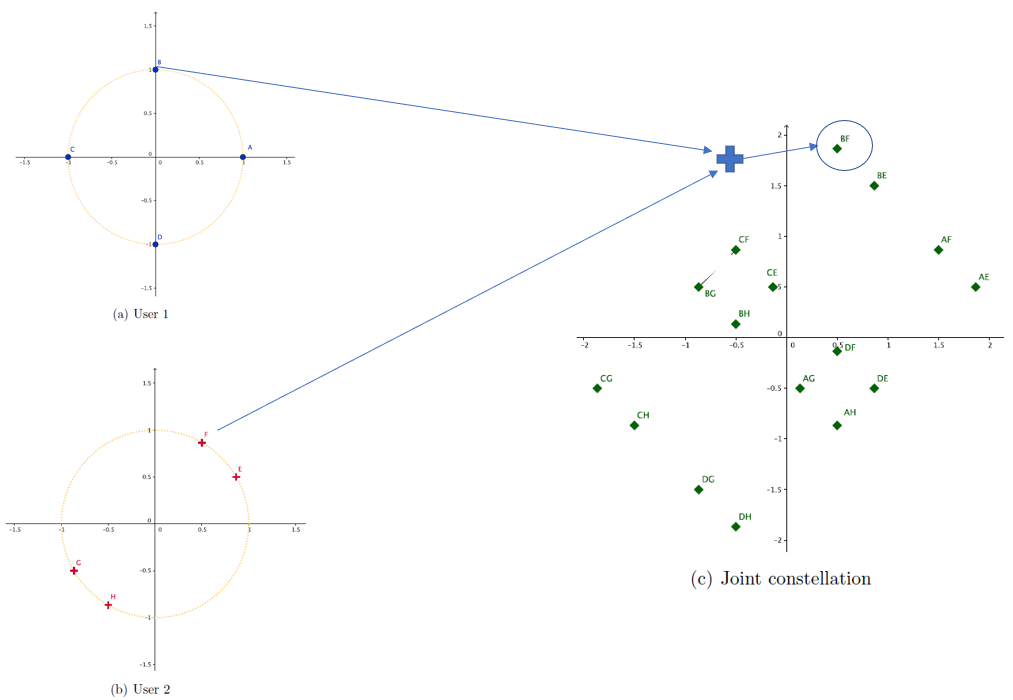}
\caption{UEP Design for 2 users with QPSK}
\label{uep}
\end{figure}

An example of the constellation for UEP criteria is shown in \reff{uep}. In this case, two users access the medium; user 1 uses the constellation shown in \reff{uep}(a), while user 2 uses the one shown in \reff{uep}(b). We can see how the symbols for user 2 maintain different distances among them and another separation compared to user 1. 
The joint constellation for the UEP case is shown in \reff{uep}(c). For example, the joint symbol marked as BF is the sum of the symbol B belonging to user 1 and the symbol F belonging to user 2.

\begin{figure}[!ht]
\centering
\includegraphics[width=1\columnwidth]{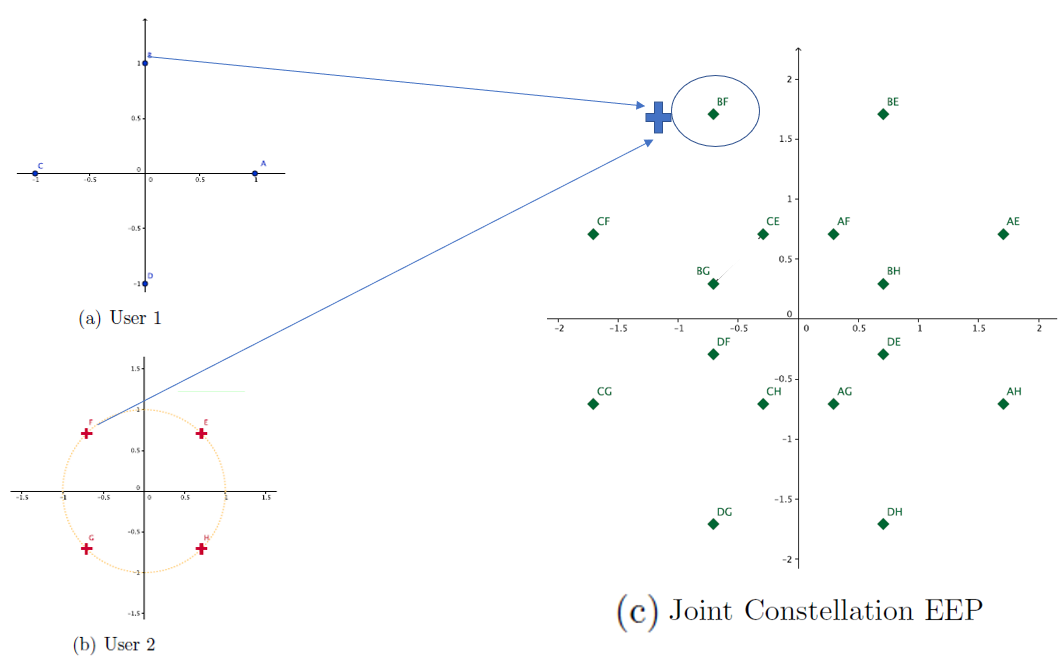}
\caption{EEP Design for 2 users with QPSK}
\label{eep}
\end{figure}

\reff{eep} shows an example of design under the criterion that all users obtain the same performance (EEP). In this case, the constellation shown in \reff{eep}(a) is for user 1, while the constellation shown in \reff{eep}(b) is for user 2. To achieve the same performance, constellation 2 is a 45-degree rotated version of user 1. The jpint constellation for this EEP example is shown in \reff{eep}(c). 

The joint symbols of this constellation are formed by combining all the symbols of the individual constellations. In both UEP and EEP cases, the rest of joint symbols are marked in the figures using letters to identify the individual symbols that form the joint symbol. 

\begin{figure}[!ht]
\centering
\includegraphics[width=1\columnwidth]{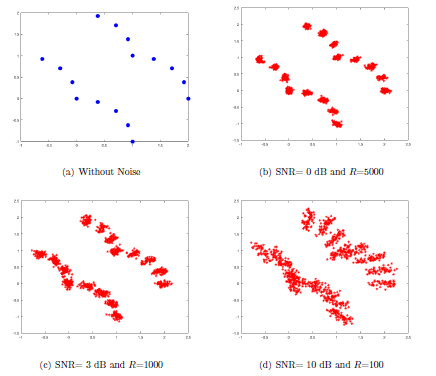}
\caption{Multiple Access using Constellation in the receiver.}
\label{noise}
\end{figure}

As mentioned, from the receiver point of view, the received joint constellation considering noise is shown in \reff{noise}. Considering massive MIMO, the number of antennas ($R$) allows us to concentrate the point cloud to reduce noise, equivalent to operating in a low signal-to-noise ratio (SNR) regime.

\section{Hybrid Multiple Access}
\label{sec:hybrid}

\begin{figure}[!ht]
\centering
\includegraphics[width=1\columnwidth]{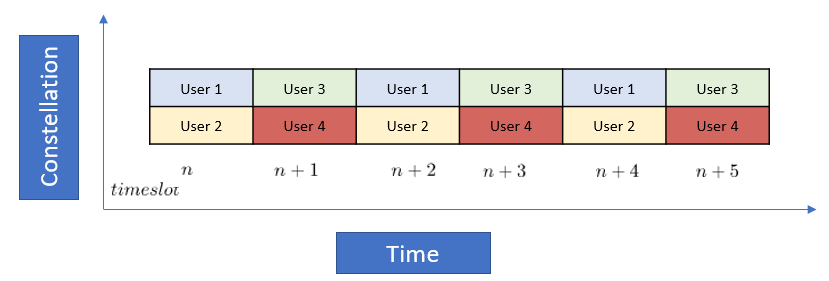}
\caption{Hybrid Multiple Access: Constellation vs Time Domain.}
\label{hibrido}
\end{figure}

Multiple users can access the system by combining the constellation domain with other traditional orthogonal resources such as time, frequency or codes to achieve the maximum possible capacity and consider all possible resources. For example, we consider a case with 4 users who are grouped into two groups of two users each group. Within each group, each of the users uses a different constellation to access non-orthogonally the system resources by non-coherent massive MIMO, while each group is temporarily separated using a TDMA scheme. To illustrate the hybrid model, we will consider the temporary resource as an example, equivalent to other resources such as frequency or codes. This combination is the hybrid scheme shown in the \reff{hibrido}.

\begin{figure}[!ht]
\centering
\includegraphics[width=1\columnwidth]{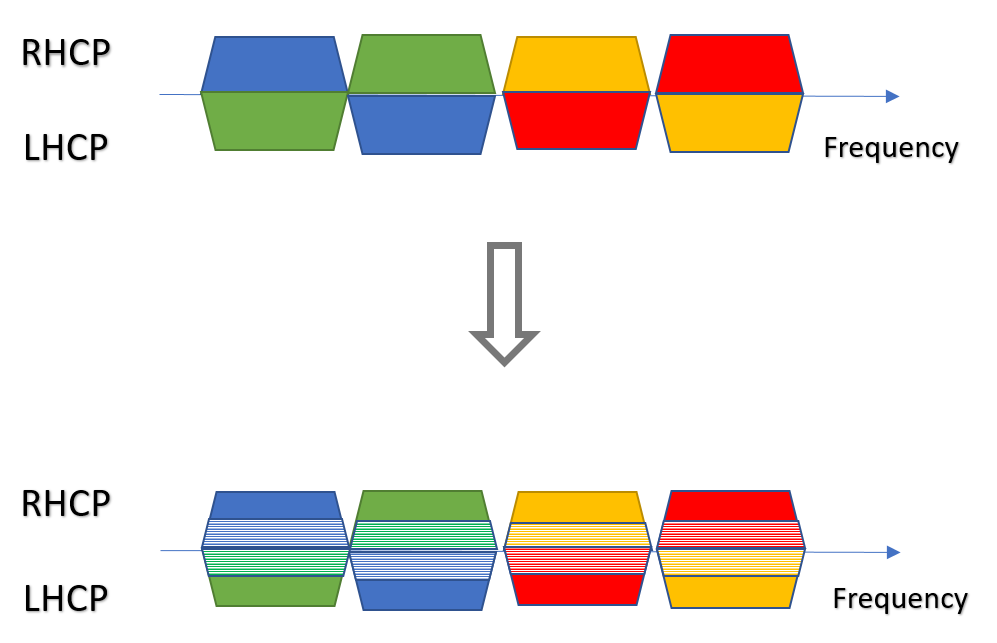}
\caption{New frequency plan.}
\label{plan}
\end{figure}

In satellite communication, the hybrid model serves to increase the capacity of the system for example in High Throughput System (HTS) satellites. The frequency plan, using a classic four-colour scheme, is updated as shown in \reff{plan}. This figure shows that the bandwidth per beam and total is maintained. Each of the beams used doubles its capacity by using the constellation domain. In a classical scheme, a color represents a frequency and a polarization (RHCP or LHCP) represented in the upper part of the \reff{plan}. In the case of adding the constellation as a new domain, we now have a color that represents the triplet: frequency, polarization and constellation. The representation of the new frequency plan to indicate the increased capacity is shown at the bottom of \reff{plan}. 

\section{Scenarios for Satellite Communications (SatCom)}
\label{sec:op}

To incorporate a non-coherent massive MIMO-based scheme in a satellite communications scenario, we have two options: 

\begin{itemize}
    \item Uplink Scheme: where non-coherent enabled massive MIMO receiver is deployed at the satellite payload.
	\item Downlink Scheme: where non-coherent enabled massive MIMO receivers are implemented at the ground segment. 
\end{itemize}

We propose three scenarios that can take advantage of the advantages of non-coherent schemes for multiple access shown in \reff{scenarios} and \reff{terrestre}. 

\begin{enumerate}
    \item \textbf{VSAT Networks (Uplink)}: this scenario includes geostationary satellites (GEO) and very small access terminals (VSAT). Multiple access technique using the constellation is applied to uplink between the VSAT terminals and the satellite. We consider the GEO satellite a high throughput multibeam satellite where the terminals are covered by the same beam access using different constellation schemes. A group of possible constellations can be defined for each beam. On the other hand, in the uplink between the hub terminal and the GEO satellite, the hubs also access the satellite using different constellation schemes.
    
    \item \textbf{Mega-constellations}: We propose to use multiple access techniques in the constellation for communications using the emerging mega-constellations formed by low orbit (LEO) or medium orbit (MEO) satellites. In this scenario we have several cases:
    
    \begin{enumerate}
        \item Gateway stations (GW) on the ground segment access the satellite on the uplink, with each GW using a different constellation. The set of possible constellations is defined for each beam. In this case, a GW station can receive signals from multiple LEO satellites if we consider site diversity. In this way, several adjacent satellites can access a GW simultaneously, each using a different constellation scheme for the downlink. 
        
        \item On the user side, for example, to offer service to a maritime fleet, ships covered by the same beam can access the LEO satellite using multiple access in the constellation domain. In the same way as the GW, a boat can receive signals from different satellites, each of them modulated with a different constellation scheme. This technique also helps the positioning of ships.
    \end{enumerate} 
    
    \item \textbf{Terrestrial-NTN}: In a scenario in which the satellite is already considered a member of the terrestrial networks as foreseen for 6G, the terrestrial users of the NTNs are separated or access the BS using different constellations between them. The same beam serves both users. In this case, a LEO satellite can be used the same way as GEO.
\end{enumerate}



\section{Challenges}
\label{sec:ch}

The proposed scenarios entail a series of challenges to be solved: 
\begin{itemize}
    \item User scheduling within each beam must be studied to optimize the system's capacity under heterogeneous channel conditions \cite{grouping}.
    
    \item The inter-user interference complicates their usage in Rician channels. Techniques to compensate for the component Line of Sight (LOS) should be considered. Different channel models, as those shown in \cite{channel} have to be considered for non-coherent schemes.
    
    \item Set of constellations optimized to reduce the Peak Average to Power Ratio (PAPR), which does not saturate the satellite's payload.
       
    \item For the hybrid model, find the optimal threshold of how many users can be multiplexed in the constellation and how many in a traditional domain, such as time, so the total system capacity is maximized \cite{15}.  
    
    \item Minimize the number of antennas on the receiving side. Above all, this is important when the massive MIMO system is located on the satellite.

    \item New scenarios for space-air environment as UAV or HAPS \cite{hap}.
\end{itemize}
   
\section{Conclusions}
\label{sec:conclusions}

In this paper, we propose to use non-coherent massive MIMO schemes as a multiple access technique. Each user accesses the system using a different constellation scheme. These constellation designs are based on DPSK under two criteria: ensuring that all users acquire the same level of performance or different classes and promoting different qualities of service. Satellite communications have presented various scenarios for this new model of accessing the medium simultaneously between users. These techniques resolve the orthogonal resource shortage to integrate the satellite into future 6G networks. 




\end{document}